\documentclass[prb,twocolumn,showpacs,superscriptaddress,floatfix]{revtex4}
\usepackage{graphicx}
\usepackage{dcolumn}
\usepackage{bm}
\begin{document}

\title{Mechanical versus thermodynamical melting in pressure--induced
amorphization: the role of defects}
\author{S. Bustingorry}
\affiliation{Centro At\'omico Bariloche and Instituto Balseiro,
Comisi\'on Nacional de Energ\'{\i}a At\'omica,
(8400) Bariloche, Argentina}
\author{E. A. Jagla}
\affiliation{The Abdus Salam ICTP, Strada Costiera 11, (34014) Trieste, Italy}
\date{\today}

\begin{abstract}
We study numerically an atomistic model  which is shown to 
exhibit a one--step crystal--to--amorphous transition upon 
decompression. The amorphous phase cannot be distinguished
from the one obtained by quenching from the melt.
For a perfectly crystalline starting sample, the transition occurs at a pressure at which 
a shear phonon mode destabilizes, and triggers a cascade 
process leading to the amorphous state. When defects are present,
the nucleation barrier is greatly reduced and the transformation
occurs very close to the extrapolation of the melting line to low temperatures.
In this last case, the transition is not anticipated by the softening of
any phonon mode.
Our observations reconcile different claims in the literature 
about the underlying mechanism of pressure amorphization.
\end{abstract}

\maketitle

\section {Introduction}

Since the first reports on pressure--induced amorphization (PIA), 
there has been an increasing interest in the phenomenon that
has been observed in a large class of systems,  
both upon compression and decompression (see Ref. [1] for a
review and references on particular materials).
The experimental interest stems from the fact that PIA occurs 
in some extremely widespread and important materials, namely water
\cite{mishima}
and quartz and its polymorphs \cite{silica}, 
and also because PIA
provides a novel route to the synthesis of amorphous materials, in
addition to the traditional technique of quenching from the melt.
>From a basic point of view, there are controversies about the mechanism
of PIA and the nature of the amorphous phase, in particular
regarding whether it is 
equivalent or not to the one obtained by quenching \cite{sio2anis,iniguez}.
Since in most cases crystalline phases that undergo PIA show reentrant
melting in the $P$--$T$ plane, it has been suggested 
\cite{mishima} that PIA is nothing but melting at temperatures
below the glass temperature of the supercooled fluid.
Other studies (specially numerical ones
\cite{softening1,softening2,grande,sciortino})
have instead emphasized the
relationship between PIA and mechanical instabilities.
It has been observed in fact, that in many cases PIA is triggered by the 
softening of a shear phonon mode \cite{binggeli94}.
In addition, some memory effects \cite{memory} and anisotropic properties
\cite{sio2anis}, although controversial \cite{contro},
show that many times what appears to be an amorphous phase preserves
within its structure signatures of the parent crystalline phase.

In view of the broad phenomenology briefly stated above,
it is highly desirable to take advantage of model systems 
in which amorphization can be studied in a transparent way, allowing
to look in detail into the mechanisms and characteristics of the transformation.
This motivates the present work, 
in which we study a simple
two--dimensional system of identical point--like particles interacting through a
specially devised two--body potential. 
The simplicity of the model allows us to
study large systems with some amount of defects, 
and observe directly the crucial role they play in the transformation.

In the model we study, there are a few different crystalline ground states
depending on the applied pressure. We present here the results of 
the evolution of the most compact structure
(stable at the highest pressures) upon
pressure release \cite{decompress}.
As we will see, PIA at temperatures in which particle diffusion is negligible 
is always related to mechanical instabilities. For perfect lattices, PIA is
reflected in the softening of a shear phonon mode. This leads to local
distortions that produce the destabilization of new vibrational modes, leading
to a cascade of instabilities \cite{grande}, which drives the system toward
an amorphous structure.
However, in the presence of defects, \textit{localized} vibrational 
modes exist, that may become unstable before 
any extended vibrational mode does.
This favors the nucleation of the amorphous
phase at pressures much closer to the
thermodynamic equilibrium value between the crystalline and amorphous phases. 

The work is divided as follows. In Section II we present the model. The results at 
zero temperature are contained in Section III. In Section IV we give evidence that the
disordered samples obtained can be called trully amorphous. In Section V we present results
at finite temperatures, and Section VI contains some discussion and conclusions.

\section{The model}

\begin{figure}[!tbp]
\includegraphics[width=8.5cm,clip=true]{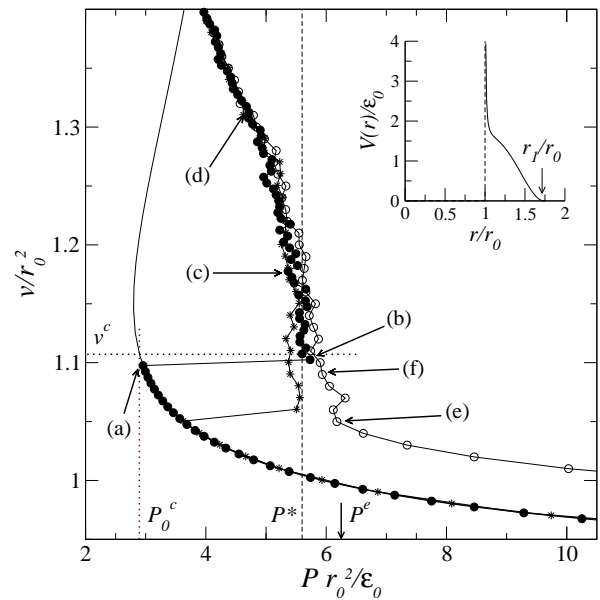}
\caption{\label{f1}
Evolution of pressure $P$ upon increasing of the specific volume $v$,
at zero temperature. 
The thin continuous line is the expected evolution if the system remains
always triangular.
The dotted lines marks the ideal stability limit $(v^c,P^c)$, at which a phonon
energy vanishes.
Dots indicate results of simulation for a perfect lattice (full circles),
a lattice with a single vacancy (stars) and a system with grain boundaries
(open circles, see Fig. \ref{f2}). 
Letters correspond to snapshots in Fig. \ref{f2}.
The inset shows the interparticle potential $V(r)$.}
\end{figure}

An isotropic, purely repulsive interparticle potential is used, with a strict
hard core at a distance $r_0$ plus an almost triangular repulsive shoulder.
The pairwise interaction potential $V(r)$ between two particles separated a 
distance $r$ is given by \cite{jagla2001}
\begin{eqnarray}
\label{ur}
V(r)&=&\infty \text{ for } r<r_0
\nonumber \\
V(r)&=&\varepsilon_0\left[1.2-2.8125 (r/r_0-1.08)^2+\frac{0.008}{r/r_0-1}\right]^2 
\nonumber \\
& &\text{ for } r_0<r<r_1\\
V(r)&=&0\text{     for } r>r_1  \nonumber
\end{eqnarray}
where $r_1=1.17315 r_0$, and $r_0$ sets the length unit.
The potential is plotted in the inset of Fig. \ref{f1}.
This kind of potential has been previously used
to systematize the anomalous properties of tetrahedrally coordinated
materials \cite{jagla98,jagla2001,sadr}. 
Then we expect they are also appropriate
to study amorphization under pressure, since this phenomenon occurs for most
of these materials.
Its crucial characteristic is the existence of two possible 
equilibrium distances between particles.

The system is simulated by standard molecular dynamics in the NVT ensemble with
periodic boundary conditions. At a given volume, the quantities of interest are
evaluated, and the volume is changed in steps of the order of $0.01\%$
by rescaling all coordinates of the particles and size of the simulation box.
Temperature is fixed by rescaling the velocities of the particles whenever necessary.
As the amorphization process implies the existence of mechanical instabilities, we typically
observe that kinetic energy tends to increase during amorphization. In a real situation
this energy transforms into heat. Here we simply eliminate it by the mentioned rescaling procedure.
Pressure is calculated by a
direct evaluation in terms of the interparticle forces. We choose
to model the system at constant volume in order to survey all regions
of the volume--pressure curve, including those 
that would be unstable in constant pressure simulations.
The results to be presented correspond to a two--dimensional system, to facilitate
a direct visualization of the particle configurations. 
We should mention however, that the
same phenomenology was observed in three--dimensional samples with the same 
interparticle potential.

\section{Zero temperature results}

\begin{figure}[!tbp]
\includegraphics[angle=-90,width=8.5cm,clip=true]{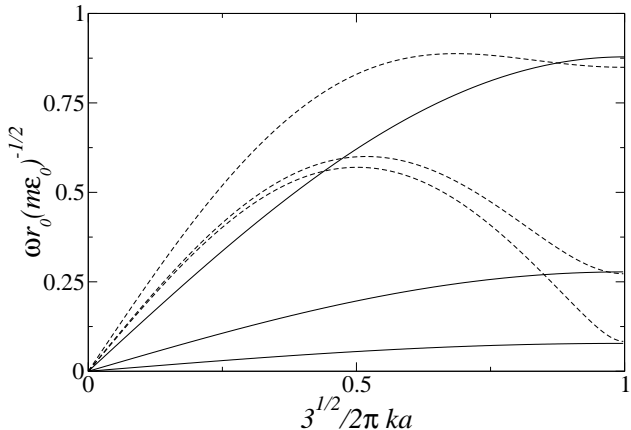}
\caption{\label{f5}
The dispersion relation of phonons along one of the directions perpendicular to
the densest planes of the triangular structure close to the instability
pressure.
Continuous lines correspond to the repulsive potential Eq. (\ref{ur}), 
and dashed lines to the same potential plus an  attractive term (Eq. (\ref{uat})) 
that includes second nearest neighbors interactions.
Curves correspond, from top to bottom, to 
$|P-P^c_0|/(r^2_0/\varepsilon_0)=0.01$, $0.001$ and $0.0001$. 
Note that since the repulsive interaction 
reaches only the first neighbors, the full branch has an analytical form of the 
type $\sim \sin(k)$, and all phonon vanish at the same pressure $P=P^c_0$.
For the case with second nearest neighbors interaction the instability occurs at
the zone border phonon $ka=2 \pi /\sqrt{3}.$
}

\end{figure}

We take as the starting configuration the one corresponding to the triangular
lattice with lattice parameter $\sim r_0$, which is stable at high pressures. 
In Fig. \ref{f1} we present the results of simulations at $T=0$ in a system of
2800 particles. 
The thin continuous line corresponds to the $v$--$P$ relation assuming the system
remains always triangular.
The reentrance of this line is a consequence of the particular form of the
interaction potential.
The numerical results for a perfect lattice (full circles) follow this line
up to some maximum volume, at which they abruptly depart from it. 
This is the pressure in which a phonon with vanishing energy first appears in 
the triangular structure (see Fig. \ref{f5}). 
Analytical evaluation shows that this instability
occurs when $3V'(a)/a+V''(a)=0$, where primes denotes derivatives of the potential, and $a$
is the lattice parameter. This expression is valid as long as the interaction between
next nearest neighbors is zero, as it is in the present case.
The critical volume $v^c$ and pressure $P^c_0$
are indicated by dotted lines 
in Fig. \ref{f1}, and they are
fully compatible with the numerical results.
The unstable phonons turn out to be shear phonons with the wave vector ${\bf k}$ 
oriented perpendicularly to one of the three most compact directions 
in the lattice.
Note that for the interparticle potential we use, the longitudinal phonon branch
along these directions is of the type $\sim \sin(k)$, and 
all shear phonons  --irrespective of the absolute value of $k$-- become 
zero energy at the same point. 
Then, in the present case the instability pressure is 
macroscopically signaled by the vanishing of the stress dependent 
shear modulus of the material \cite{gprimap,zhou96}, $\mu=(C_{11}-C_{22})/2-P=0$.

\begin{figure}[!tbp]
\includegraphics[width=8.5cm,clip=true]{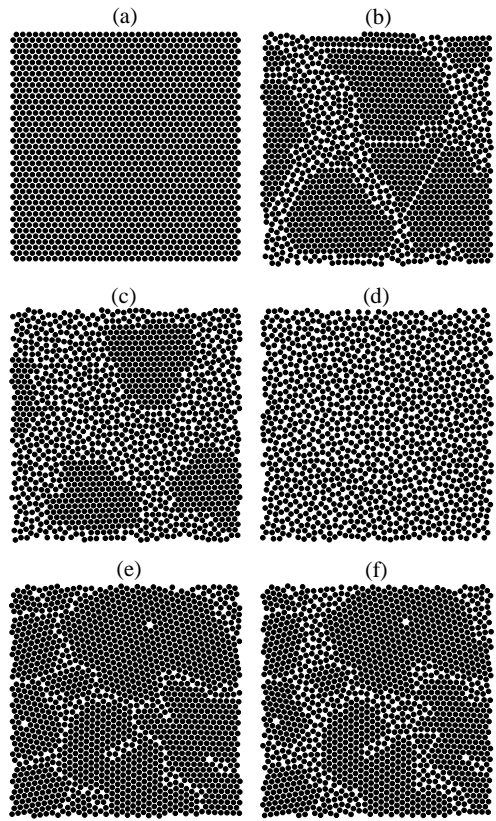}
\caption{\label{f2}
Snapshots of the systems at the points indicated 
in Fig. \ref{f1} (only about a quarter of the full simulated system is shown). 
Upper four panels correspond to a 
perfect crystal starting sample, whereas the two lower ones
are from a poly--crystalline sample with grain boundaries and vacancies.
}
\end{figure}

At $v^c$ there is a sharp and abrupt increase of pressure \cite{singular}.
Snapshots of the particle configurations give clues of what happens in the
system.
In Fig. \ref{f2}(a) we see the triangular configuration just before the 
instability.
Figure \ref{f2}(b) shows that after changing the volume a very small quantity, 
an instability has propagated in the system. 
We stress that the evolution of the system from the configuration in 
Fig. \ref{f2}(a) to that in \ref{f2}(b) is triggered by a very small volume
change, and is just the steepest descend evolution toward a local energy 
minimum of the energy landscape. In this evolution, pressure
recovers to a higher value indicated as $P^*$ (dashed line) in Fig. \ref{f1}. This value 
is roughly maintained \cite{tbp} upon further volume increase 
up to $v/r_0^2\sim 1.27$, which is close to the
volume at which the disordered regions have taken over the whole system. 
If volume is increased further, pressure decreases noticeably.

The thermodynamic equilibrium pressure $P^e$ at $T=0$
between the crystalline and disordered structures is indicated by the 
arrow in Fig. \ref{f1}.
The value of $P^e$ was calculated by enthalpy evaluations of the triangular
lattice and the amorphous structure obtained by quenching from the melt.
Thus, $P^e$ represents the natural extrapolation of the melting line to zero 
temperature. We see that $P^e$ is close to (though a bit higher than) $P^*$. 

It must be noticed that constant pressure simulations would have shown
at $P^c_0$ an abrupt transition between a completely ordered and a 
completely disordered system, while $P^*$ would have remained hidden.
However, $P^{*}$ has a clear physical meaning as the transition pressure
\textit{once the disordered phase has been nucleated}.
This suggests that if nucleation centers are present in the system, $P^{*}$ 
will be experimentally accessible as the actual transition pressure.
In fact, Fig. \ref{f1} shows also the $v$--$P$ evolution for a system with a
single vacancy and for a system with grain boundaries. 
In the former case, the 
instability occurs before $v^c$ is reached, and in the latter case it even
occurs without any pressure reentrance \cite{nota}. 
Note that the three curves tend to coincide
after the first destabilization of the original lattice.
Two snapshots of the system for the poly--crystalline case 
[Fig. \ref{f2}(e) and (f)] show how the disordered phase grows from grain
boundaries, that provide nucleation centers for the transformation.
These results are a clear indication that defects are very effective in 
lowering the energy barrier for the transition, and also explain why, in 
experiments, transformation pressures close to the thermodynamic values are 
usually observed.
The lowering of the energy barrier due to defects and its relation to
thermodynamic melting has been discussed by Mizushima \textit{et al.} 
\cite{mizu94} in the context of crystal--crystal 
pressure--induced transformations.
Note that in the case in which the transformation is triggered by defects, the phonon
spectrum of the system (and in particular the elastic constants) gives no indication of the
instability that is about to occur. This is particularly obvious in the case of a single
vacancy: a single defect cannot modify the phonon spectrum of an infinite sample,
and then its effect it is not seen in the elastic constants, but it produces a finite change 
in the value of the critical pressure $P^c_0$\cite{single}.

It was already mentioned that for the interaction potential used, 
the whole branch of shear phonons become zero energy
at $P^c_0$. We want to emphasize however that amorphization it is not related to 
this degeneracy, 
as it can also be triggered by a single phonon becoming unstable. 
In fact, we did simulations with a modified potential in which
an attractive term $V_a$ reaching up to second nearest neighbors was included. 
The actual potential
used is the one of the previous section plus an attractive term given by
\begin{eqnarray}
V_a(r)&=&2 \varepsilon_0 \left(r-r_1\right)^2 \text{ for } r_1<r<\left(r_1+2r_0\right)/2,
\nonumber \\*
V_a(r)&=&-2 \varepsilon_0 \left[\left(r-2r_0\right)^2-\left(r_1-2r_0\right)^2/2\right] 
\nonumber \\*
& &\text{ for } \left(r_1+2r_0\right)/2<r<2r_0.
\label{uat}
\end{eqnarray}
The instability corresponds now to the vanishing of the energy of edge zone shear
phonons only, as indicated in Fig. \ref{f5}.
The amorphization of the system occurs precisely at the point where this phonon 
becomes zero energy, and it was observed
to be similar to the previous case where the full shear phonon
branch becomes unstable at the same pressure.
This rules out an amorphization mechanism
in which at the instability, arbitrary combinations of the unstable phonons 
generate disordered movements of the atoms, and favors a cascade 
mechanism as described in the discussion section below.


\section{Comparison between decompressed and quenched amorphous}

\begin{figure}[!tbp]
\includegraphics[angle=-90,width=8.5cm,clip=true]{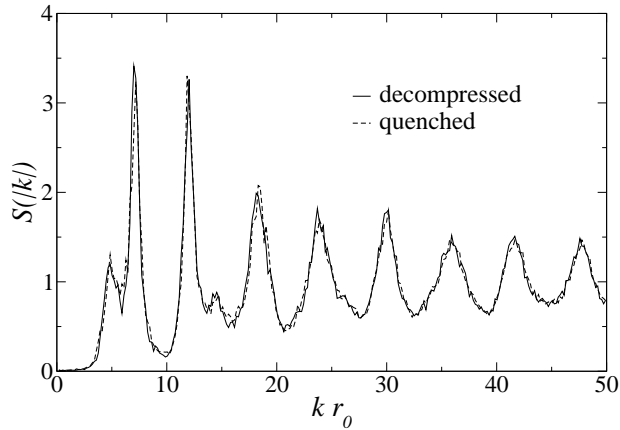}
\caption{\label{f3}
The structure factor of the amorphous obtained by decompression of
a perfect crystalline sample and that obtained by a quench from the liquid
phase, at the same volume $v/r_0^2=1.289$.
They coincide within the numerical precision.
}
\end{figure}

The claim that the disordered structure we obtain upon decompression is 
really amorphous is supported by the following facts.
We find no evidence of orientational order in the structure factor $S({\bf k})$,
even in the case in which we start with a mono--crystalline sample.
Moreover, a quantitative comparison with the structure factor of a system
quenched from the liquid (prototypic of an amorphous sample)
shows that they are indistinguishable (Fig. \ref{f3}).
In addition, we have failed to find any
single systematic difference between the two amorphous.

\section{Finite temperature effects}\label{V}

If temperature is not strictly zero, even the perfect lattice
amorphizes before the ultimate mechanical instability limit is reached.
Simulations at finite temperature with perfect samples show  that the amorphization
occurs at a temperature dependent critical pressure $P^c(T)$ (such that $P^c(T=0)=P^c_0$). 
The temperature dependence of the critical pressure 
originates in the fact that a thermally activated process may destabilize the soft 
phonon before it actually becomes zero energy.

It is instructive to see in some more detail how this destabilization occurs.
Consider the normal modes (phonons) of the system. We will study the case in which
only a single phonon with wave vector ${\bf k}$ perpendicular to one of the high density planes
has non-zero amplitude.
We expand the energy of the lattice in powers of the amplitude of the phonon oscillation, going 
one order beyond the harmonic approximation.
Let us call $\delta_l$ and $\delta_t$ the longitudinal and transversal
oscillation amplitude of that phonon.
As we will see there is a nontrivial coupling between the two at the instability.

\begin{figure}[!tbp]
\includegraphics[angle=-90,width=8.5cm,clip=true]{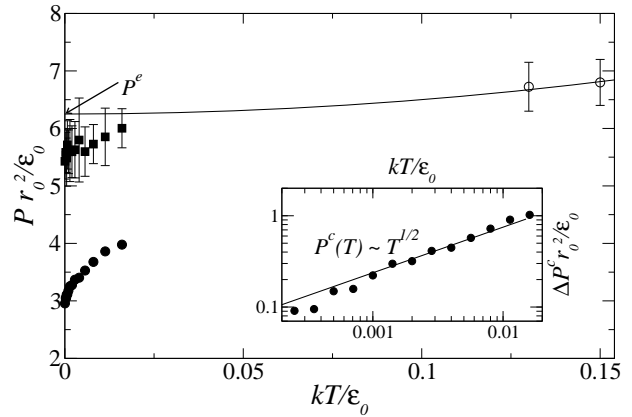}
\caption{\label{f4}
Evolution of $P^c$ (full circles) and $P^{*}$ (full squares) as a function of 
temperature (bars in $P^*$ indicate the whole range in which transformation 
occurs).
The inset shows in a log--log plot that $\Delta P^c=P^c(T)-P^c(T=0)$ increases 
as $T^{1/2}$ at finite temperatures.
The extrapolation of the melting line (from calculations at higher temperatures, two points
are seen)
is indicated by the dashed line. It goes to $P^e$ at $T=0$.
}
\end{figure}

\begin{figure}[!tbp]
\includegraphics[angle=-90,width=8.5cm,clip=true]{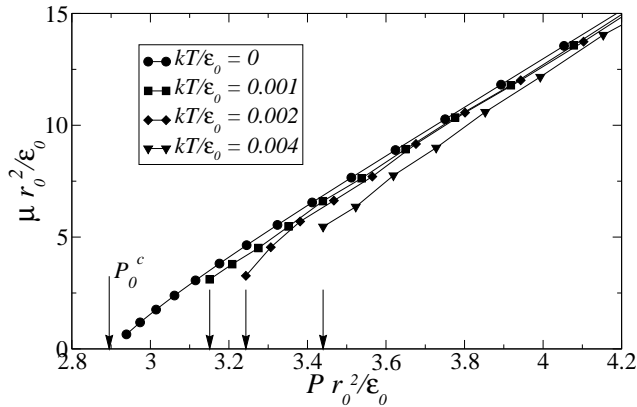}
\caption{\label{f6}
Variation of the stress dependent shear modulus $\mu$ with pressure at different
temperatures. For $T=0$, $\mu$ vanishes at the critical pressure $P^c(T=0)$, but
for increasing temperatures $\mu$ does not tend to zero at the corresponding
critical pressure $P^c(T)$, showing that the transition it is not
macroscopically captured by the elastic constants. The numerical values of 
$P^c(T)$ are indicated by the vertical arrows.
}
\end{figure}

Taking into account that our potential produces interactions only among first neighbors in the lattice,
the energy per particle $e_0$ of the system up to third order in the amplitudes can be written
after some lengthy but straightforward calculation as

\begin{eqnarray}
\label{e0}
e_0&=&\frac{1}{8}\sin^2\left(\frac{\sqrt{3}ka}{4}\right) \nonumber \\
& &\times \left[ \delta_t^2\left(\frac{3V'}{a}+V''\right)  
+\delta_l^2(\frac{V'}{a}+3V'')\right] \\
& &+\frac{\sqrt{3}}{8}\sin^3\left(\frac{\sqrt{3}ka}{4}\right) \left[
\delta_t^2 \delta_l  \left(-\frac{4V'}{a^2}+\frac{V''}{a}+V'''\right)\right].
\nonumber
\end{eqnarray}
The first line in (\ref{e0}) is the harmonic contribution, the
second is a cubic term in the displacements (there is an additional cubic term proportional
to $\delta_l^3$, but this can be shown to be not relevant for the analysis below). 
The system becomes unstable when the
coefficient of $\delta_t^2$ vanishes, providing again the condition 
$3V'/a+V''=0$, which
defines the $T=0$ values $P^c_0$ and $v^c$. Close enough to this instability point, 
$3V'/a+V''$ is proportional to $P-P^c_0$ and to lowest order all other
coefficients can be considered to be constants. Then generically, the energy can
be written as

\begin{eqnarray}
e_0&=&\sin^2\left(\frac{\sqrt{3}ka}{4}\right) \left[ A(P-P^c)\delta_t^2  +B\delta_l^2\right] \nonumber \\
&+&\sin^3\left(\frac{\sqrt{3}ka}{4}\right) \left[D\delta_t^2 \delta_l  \right].
\label{dos}
\end{eqnarray}
The existence of a $\delta_t^2 \delta_l$ term is very important, 
as it indicates that  for $P$ slightly larger than
$P^c_0$ there is a saddle point (actually two, one with positive and the other
with negative $\delta_t$) close to $\delta_t=0$, $\delta_l=0$, than can be
determined requiring stationarity of (\ref{dos}). The result is

\begin{eqnarray}
\delta_t^S&=&\pm \frac{\sqrt{2AB(P-P^c)}}{D\sin(\frac{\sqrt{3}ka}{4})},
\nonumber\\
\delta_l^S&=&\frac{A(P-P^c)}{D\sin(\frac{\sqrt{3}ka}{4})},
\label{tres}
\end{eqnarray}
where the upperscript indicates the values at the saddle. 
If the system reaches this saddle it can escape from the local minimum 
at the origin, namely, the system destabilizes. The energy
barrier $h$ for this process is obtained by reinserting (\ref{tres}) into (\ref{dos}). The result
is:
\begin{equation}
h=\frac{4A^2B(P-P^c)^2}{D^2}.
\end{equation}
Note that the barrier is exactly the same whatever the wave vector of the phonon
considered
(this is no longer true if further neighbors interactions are included, 
but the possibility to escape through the jump of a barrier remains). 
The instability mechanism is then driven by the $\delta_t^2 \delta_l$ term in the energy,
what implies a not trivial coupling of transverse modes (the one actually having vanishing frequency) 
and longitudinal ones. At finite temperatures the barrier $h$ can be surmounted. Nucleation theory
tells that escape time $\tau$ is proportional to $\exp {(h/T)}$. Assuming the preexponential factor
is a constant $\tau_0$ (this is certainly not true, but dependences on $\tau_0$ become weak in the final result)
we obtain that if a time $t_0$ is given, the system will overcame the
energy barrier if $t_0\gtrsim \tau$, and from here
we obtain the formula
for the temperature necessary to escape a given barrier, namely:
\begin{equation}
T>h \ln ^{-1}(t_0/\tau_0).
\end{equation}
For practical purposes the logarithmic factor can be usually taken to be approximately 0.1.
As $h\sim (P-P^c_0)^2$,
we obtain that 
the critical pressure $P^c(T)$ increases as $T^{1/2}$ at finite temperatures, 
a behavior that is well reproduced in the simulations (Fig. \ref{f4}).
The $P^c(T)$ line is then the pressure amorphization line for a sample without defects,
and can be properly interpreted as the `mechanical melting'
line of this system. In the absence of defects it marks the limit on which
the crystalline phase destabilizes (purely mechanically at $T=0$, and by a thermally activated
process at $T\ne0$). 
Note however, that the elastic constants of the system, in particular
the shear modulus $\mu$, do not extrapolate to zero at $P^c(T)$, except at $T=0$ (see Fig. \ref{f6}). 
In fact, the numerical
evaluation by means of fluctuation formulas \cite{zhou96,constelast} 
shows that the effect of temperature in the elastic constants is very
small. At $T=0$ the stress dependent shear modulus $\mu$ vanishes at $P^c_0$, as we already knew from the
phonon dispersion relation, but at $T\ne0$ extrapolation is consistent with a vanishing of $\mu$
very close to $P^c_0$, and certainly not at $P^c(T)$. The thermally
activated process that at finite temperatures leads to the instability at the
higher pressure $P^c(T)$ is a rare
event that it is enough to occur once to completely destabilize the system. This process is
not captured in the value of the elastic constants. 

Up to here the results for samples without defects.
For defective samples the amorphization line $P^{*}(T)$ is only weakly dependent on $T$. 
In addition, it is very close to the extrapolation of 
the melting line obtained from simulations at higher temperatures (see Fig. \ref{f4}).
$P^{*}(T)$ and $P^c(T)$ are well different from each other, 
particularly at very low temperatures. Defects are responsible for this difference, and provide the
link between the `mechanical' and `thermodynamical' scenarios for pressure amorphization.

\section{Discussion and Conclusions}

The experimental observation that in some materials PIA occurs roughly along
the extrapolated melting line has motivated the suggestion
\cite{mishima} that `thermodynamic melting' is the underlying driving force
for amorphization. 
Our results show that amorphization at zero temperature is always a
phenomenon related to mechanical instabilities.
In perfect lattices, mechanical instabilities reflect necessarily in
the phonon spectrum of the material, and amorphization can be 
easily related to a mechanical process.
However, for defective lattices, the mechanical instability that triggers
amorphization is associated with localized vibrational modes around defects, 
and this instability occurs typically without any noticeably signature in the 
phonon spectrum, i.e., the mechanical nature of the process is more subtle.
We emphasize that a single defect is sufficient to trigger the amorphization 
of a whole macroscopic sample\cite{single}.

\begin{figure}[!tbp]
\includegraphics[angle=-90,width=8.5cm,clip=true]{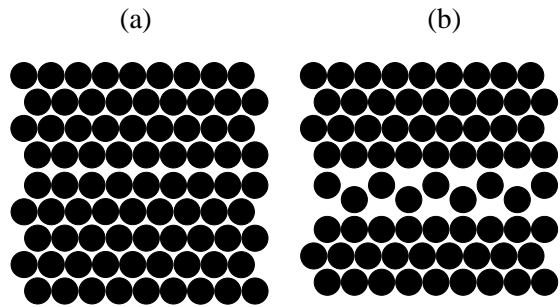}
\caption{\label{f7}
The first steps in the cascade process from a perfect 
to an amorphous lattice: The original perfect lattice becomes unstable 
when the energy of a phonon vanishes, and this drives a transition to a 
sample with a shuffle (a). 
This configuration however is not stable and spontaneously evolves 
toward the configuration in (b). 
But this is not stable yet, and the process continues until
equilibrium is reached when a finite fraction of the system 
is transformed (compare with Fig. (2) (a) and (b)).
}
\end{figure}

How this first instability is able to transform the highly ordered original lattice
into a fully amorphous configuration is not easy to understand, specially considering
that this evolution is fully deterministic from a mechanical point of view, 
and can be described as the steepest descendant path of the configuration point of the system
onto its energy landscape.
A deep understanding of this process will surely shed light also onto the very 
definition of what an amorphous material is, typically characterized by what it lacks, instead of
what it possesses.
Here we make only the following considerations \cite{grande}.
The first instability of the lattice (at $T=0$) can always be described as the
vanishing of the frequency of one of the normal modes of the system. 
This normal mode may be a phonon for a perfect lattice, or a 
localized mode for a defective lattice.
When the coordinate characterizing the destabilizing mode grows, it typically goes beyond
the applicability limit of the linear theory, and the whole mechanical stability of the system 
has to be reanalyzed. Then, although the analysis of Section \ref{V} is appropriate to determine the barrier for the
system to scape from its metastable equilibrium, it cannot be continued when the system actually
overcomes the saddle without including the interaction with other modes. 
For instance, it occurs in our system that once the coordinate of the unstable phonon starts to increase, it
couples to other phonons and the results is that a shuffle of two pieces of the material develops.
In this way, the instability generated by an extended object (the unstable phonon) produced a localized 
(one--dimensional) defect on the lattice.\cite{fonon}.
But it turns out that, in the present case,
the new configuration of the lattice is not stable either, and a new distortion 
occurs spatially close to the region
already distorted (see Fig. \ref{f7}). 
In the present case of amorphization, 
it is likely that a cascade of these processes occurs
that leads to
an amorphous structure. The spatially localized nature of the process
reflects in the fact that for partially transformed samples, 
transformed and untransformed regions are spatially 
separated, as seen in Fig. \ref{f2}.
We however emphasize that another possibility could be 
that after the first phonon instability and the first shuffle, the system reaches
a mechanically stable configuration, then further increases of the volume will produce independent shuffles in the samples, along
the three directions related by symmetry.
If this is the case, then this single phonon will drive a transformation 
(which has the character of a martensitic transformation) to a new 
(poly--)crystalline phase. This and other intermediate cases, that in our
model occur for different parameters, will be discussed separately.

In summary, we have arrived at a picture of the mechanism of PIA,
reconciling mechanical and thermodynamical melting in a unifying scheme.
The present model exhibits a one--step crystal--to--amorphous transition to a
truly amorphous state that does not differ from the disordered structure
obtained by quenching from the melt. Once the transformation is triggered, it is
followed by a cascade process that produces the amorphous
structure.
For a perfect lattice, and vanishingly small temperatures, the
transformation takes place through a shear phonon instability at a
pressure far apart the thermodynamical equilibrium pressure. For defective
lattices, the mechanical instability shows up by the destabilization of 
localized vibrational modes,  close to the
thermodynamical equilibrium pressure. In either case, the transformation is 
always triggered by mechanical instabilities. 
Temperature can produce the transition to occur
before the mechanical stability limit is reached, due to a process 
of thermally activated jump over a barrier.
In real samples with appropriate defects, experimental transition 
pressures will be seen close to the extrapolation 
of the melting line. These facts reconcile different claims in 
the literature regarding the nature of the PIA of materials.
We remark that, although we presented results for a two--dimensional model
system, preliminary results in three--dimensional samples qualitatively agree
with the phenomenology stated above, and will be presented elsewhere.
We finally stress that these considerations have potential applications in the
study of melting mechanisms of crystalline materials\cite{jin}.

\section{Acknowledgments}

We thank N. Binggeli for useful comments on the manuscript.
S. B. is financially supported by CONICET (Argentina).


\begin{thebibliography}{10}
\bibitem{decompress}Materials that amorphize upon decompression include CaSiO$_3$,
MgSiO$_3$, and stishovite, the stable high pressure form of SiO$_2$.
\bibitem{singular} We include a vanishingly small temperature in the simulations without
defects, in order to break the extreme symmetry of the problem in this case. This makes the
transition to occur slightly below $v_c$ in Fig. \ref{f1}. This
very small temperature can be considered to be zero in any other respect.
\bibitem{nota}The upward shift of the curve for a poly--crystalline sample at
large $P$ in Fig. 1 is a consequence of the fact that grain boundaries occupy
a finite fraction of the system. For a system with much larger crystallites,
this shift is negligible and the large $P$ part of the $P-v$ curve coincides with 
that of a mono--crystalline system.
\bibitem{single}In our finite samples, a `single defect' is still a large density of defects.
We are sure however that the effect of a single vacancy, for instance, does not vanish in the 
thermodynamic limit as we have verified that the change in $P^c$ produced by this single defect 
remains constant when the size of the system is doubled.
\end{thebibliography}
\end{document}